\documentclass[12pt,superscriptaddress,aps,prb]{revtex4-1}

\usepackage[hmargin=2.0cm,vmargin=2.3cm]{geometry}
\usepackage{braket}
\usepackage{textpos}
\usepackage{amssymb,amsthm}
\usepackage{amsmath}
\usepackage{hyperref}
\usepackage{txfonts}
\usepackage{color}
\usepackage{natbib}
\usepackage{etoolbox}
\usepackage{ulem}
\usepackage{graphics,epsfig,subfigure}
\usepackage{textcomp}

\newcommand{\Ca}{$^{40}$Ca$^{+}$}

\newcommand{\nbar}{\bar{n}}
\DeclareMathOperator{\tr}{tr}
\newcommand{\new}[1]{#1}
\newcommand{\old}[1]{}

\begin{document}

\title{Local probe of single phonon dynamics in warm ion crystals}

\author{A. Abdelrahman}\new{\email{Corresponding author: mabd@kth.se}}
\affiliation{Department of Physics, University of California, Berkeley, California 94720, USA}
\affiliation{Department of Physics, Stanford University, 452 Lomita Mall, Stanford, CA 94305}
\author{O. Khosravani}
\affiliation{Department of Physics, University of California, Berkeley, California 94720, USA}
\author{M. Gessner}
\affiliation{Physikalisches Institut, Albert-Ludwigs-Universit\"at Freiburg, Hermann-Herder-Str. 3, D-79104 Freiburg, Germany}
\affiliation{QSTAR, INO-CNR and LENS, Largo Enrico Fermi 2, I-50125 Firenze, Italy}
\affiliation{Istituto Nazionale di Ricerca Metrologica, Strada delle Cacce, 91, I-10135 Torino, Italy}
\author{A. Buchleitner}
\affiliation{Physikalisches Institut, Albert-Ludwigs-Universit\"at Freiburg, Hermann-Herder-Str. 3, D-79104 Freiburg, Germany}
\affiliation{Keble College, University of Oxford, OX1 3PG Oxford, United Kingdom}
\author{H.-P. Breuer}
\affiliation{Physikalisches Institut, Albert-Ludwigs-Universit\"at Freiburg, Hermann-Herder-Str. 3, D-79104 Freiburg, Germany}
\author{D. Gorman}
\affiliation{Department of Physics, University of California, Berkeley, California 94720, USA}
\author{R. Masuda}
\affiliation{Department of Physics, University of California, Berkeley, California 94720, USA}
\author{T. Pruttivarasin}
\affiliation{Department of Physics, University of California, Berkeley, California 94720, USA}
\affiliation{Department of Physics, Mahidol University, 272 Rama VI Rd., Ratchathewi, Bangkok 10400, Thailand}
\author{M. Ramm}
\affiliation{Department of Physics, University of California, Berkeley, California 94720, USA}
\author{P. Schindler}
\affiliation{Department of Physics, University of California, Berkeley, California 94720, USA}
\author{H. H\"affner}\new{\email{Corresponding author: hhaeffner@berkeley.edu}}
\affiliation{Department of Physics, University of California, Berkeley, California 94720, USA}

\date{\today}                                            

\maketitle

\textbf {
\new{
The detailed characterization of  non-trivial coherence properties of composite quantum systems of increasing size
is an indispensable prerequisite for scalable quantum computation, as well as for understanding 
of nonequilibrium many-body physics.  
Here we show how autocorrelation functions  in an interacting system of phonons
as well as the quantum discord between distinct degrees of freedoms 
can be extracted from a small controllable part of the system. As a benchmark, we show this in 
chains of up to 42 trapped ions, by tracing a single phonon excitation 
through interferometric measurements of only a single ion in the chain. We observe the spreading and partial refocusing of the excitation in the chain, even 
on a background of  
thermal excitations. We further show how this local observable reflects the dynamical evolution of quantum discord between the electronic state 
and the vibrational degrees of freedom of the probe ion.
}}

\new{\subsection*{Introduction}}
The faithful description of the state of an interacting many-particle quantum system turns ever more difficult as the system size increases\new{\cite{Chuang1997,Haeffner2005, Walschaers2016}}. Therefore, one usually focuses on collective quantifiers -- such as, e.g., the total magnetization
of a spin chain -- to distinguish macroscopically distinct many-body phases. However, engineered many-body quantum systems, such as trapped ions, cold atoms, and superconducting circuits, offer the unique and distinctive feature that their microscopic structure is accessible to experimental diagnostics \cite{Friedenauer2008a,Bakr2009,Senko2014,Hild2014,Jurcevic2015,Langen2015,Labuhn2016,Roushan2014,Hacohen-Gourgy2015}. Therefore, hitherto unaccessible dynamical features can be directly observed offering a novel opportunity to 
gain insight into the emergence of macroscopically robust features \cite{Olkiewicz2002,Buchleitner2003,Amico2008,Polkovnikov2011,Geiger2012,Gessner2014}. Further, the system size can be increased starting from only a few particles up to millions while maintaining coherent dynamics. Thus, those engineered systems allow one to study the emergence of macroscopic properties directly from the microscopic spectral and dynamical structure \cite{Senko2014,Jurcevic2015}. 

On the other hand, a microscopic approach where not only global observables but also the local ones as well as their correlations become an essential part of the description will at some point hit a complexity threshold. In particular, the increasing number of interacting degrees of freedom implies a rapidly proliferating number of possible defects, which induce disorder and/or noise; with finite temperature effects as one of the most fundamental and ubiquitous perturbations. Consequently, there is a conceptual and practical need to develop scalable and experimentally feasible approaches \cite{Knap2013,Gessner2014a,Schlawin2014,gessner2016local} to probe well-defined microscopic quantum features such as correlation functions and quantum correlations even at system sizes which prevent an 
exhaustive microscopic characterization, 
and where the presence of uncontrolled noise such as a finite temperature is unavoidable. Such methods will help to gain a better understanding of the ultimate realm of non-trivial quantum mechanical effects, on meso- and macroscopic scales. We  believe that such methods could also be useful to characterize the naturally very complex functioning of quantum computing devices and thus help push quantum computing architectures to scales which become practically relevant.

To contribute to this endeavour, we present a tagging method which can be used to track single quantum excitations in warm quantum many-body systems.
We achieve this by coupling an auxiliary qubit in form of the electronic state of a probe ion to the motion of an ion chain of variable length  \cite{DeChiara2008,Baltrusch2013,Clos2015}. We will demonstrate that this allows one to measure the autocorrelation function of the phonon dynamics even in situations where only access to a subsystem is available \cite{Gessner2014a,Schlawin2014}. Moreover, we will see that the local dynamics
of the qubit is directly linked to the dynamical evolution of the quantum discord\new{\cite{Modi2012}} between the qubit degree-of-freedom of the probe ion and its local vibrational degree of freedom. We recall  that discord describes local quantum properties of correlated states \cite{Modi2012,gessner2016local} which are relevant for quantum information tasks such as entanglement distribution \cite{Streltsov2012} and activation \cite{PhysRevLett.106.160401,PhysRevLett.106.220403}, as well as for local quantum phase estimation \cite{Girolami2014}.

\new{\subsection*{Results}}
\new{\subsubsection*{Experimental system}}
The dynamics of the transverse motion ($x,y$)  of an ion string aligned along the $z$-axis can be described by the Hamiltonian of a chain of
coupled quantum harmonic oscillators 
\citep{James1998,Ramm2014},
\begin{equation}
\label{eq:chain}
{H}_{\rm chain} = \sum_{i=1}^{N} \hbar \omega_{x,i} a_i^\dagger a_i + \hbar \sum_{i=1}^{N}
\sum_{\substack{j=1 \\ j < i}}^{N} t_{ij} \left(a_i^\dagger a_j + a_i
a_j^\dagger \right)\ ,
\end{equation}
where $a_i$ is the annihilation operator acting on the Fock state of the $i^{th}$ harmonic oscillator which represents the corresponding ion's transverse 
motional state, and the $\omega_{x,i}$ are the frequencies of the transverse motion of each ion along the $k$-vector of the qubit laser, which 
are modified by the Coulomb potential due to the interaction with neighboring ions. These frequencies are given by
$\omega_{x,i}  = \omega_x -  \sum_{\substack{j=1 \\ j \neq i}}^N t_{ij}$
in which $\omega_x$ is the transverse trap center-of-mass frequency and $t_{ij}$ is the coupling between the local modes of ions $i$ and $j$,
\begin{equation}
t_{ij} = \frac{1}{2} \frac{1}{m \omega_x} \frac{e^2}{4 \pi \epsilon_0}\frac{1}{|z_i^0-z_j^0|^3} \,
\end{equation}
with the equilibrium position  $z_i^0$ of the $i$th ion along the $z$ axis.

\begin{center}
\begin{figure}
\includegraphics[angle =0, width=0.99\textwidth]{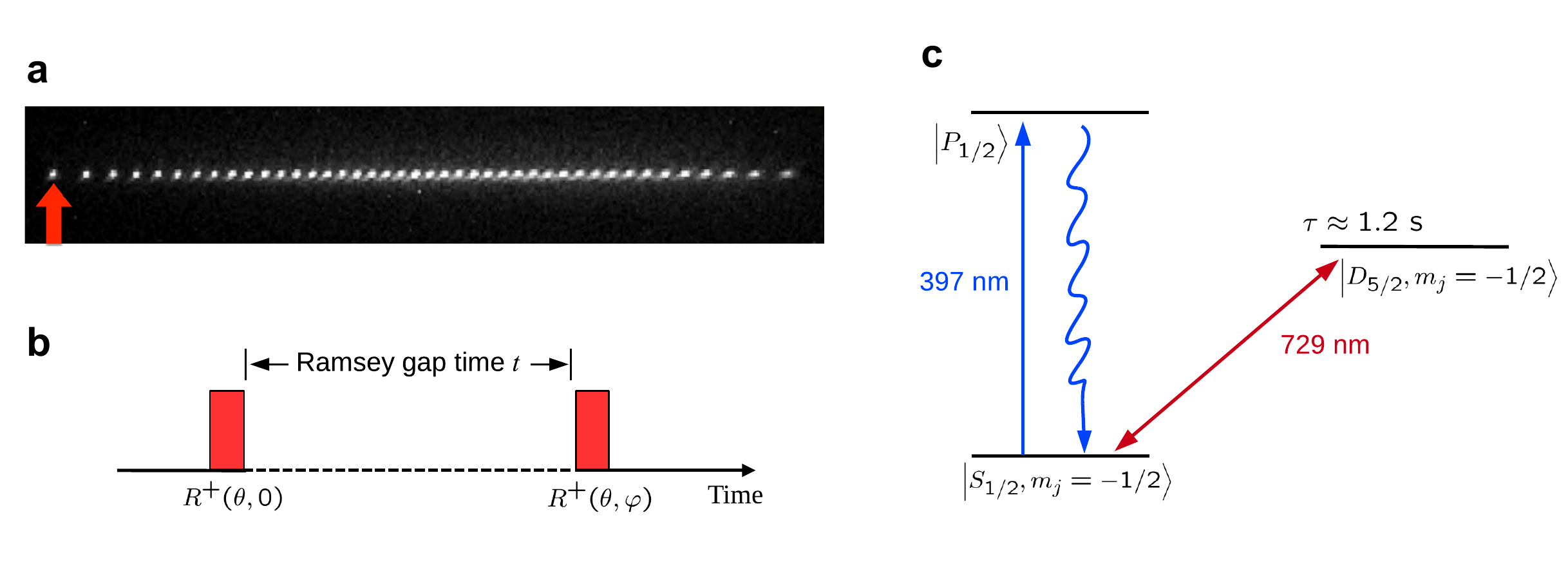}
\caption{ \label{fig:schematics}
\textbf{Schematic overview and electronic structure of \Ca.} a, image of 42 ions. The red arrow 
indicates the position of the laser beam exciting the
probe ion on the blue motional sidebands of the transverse motion with light at 729~nm. b, Ramsey sequence with a free evolution time $t$ governed by the Coulomb interaction between the ions. 
c, relevant electronic levels of \Ca. The quadrupole transition between the $\ket{S_{1/2}, m_j = -1/2}$ and $\ket{D_{5/2}, m_j = -1/2}$ states is driven by a narrow linewidth 729~nm laser beam tightly focused on the 
probe ion. The Zeeman degeneracy is lifted by applying a magnetic field of 323~{\textmu}T. Cooling and readout are performed on the $\ket{S_{1/2}}\rightarrow\ket{P_{1/2}}$ transition at 397~nm.}
\end{figure}
\end{center}

We trap strings between eight and 42 \Ca-ions in a linear Paul trap (see Fig.~\ref{fig:schematics}) with trapping 
frequencies of $\omega_x = 2\pi \times \sim 1.9$~MHz, $\omega_y = 2\pi \times \sim 2.2$~MHz and $\omega_z$ 
ranging from $2 \pi \times 94$ to $2 \pi \times 196$~kHz. 

\begin{center}
\begin{figure}
\includegraphics[angle =0, width=0.55\textwidth]{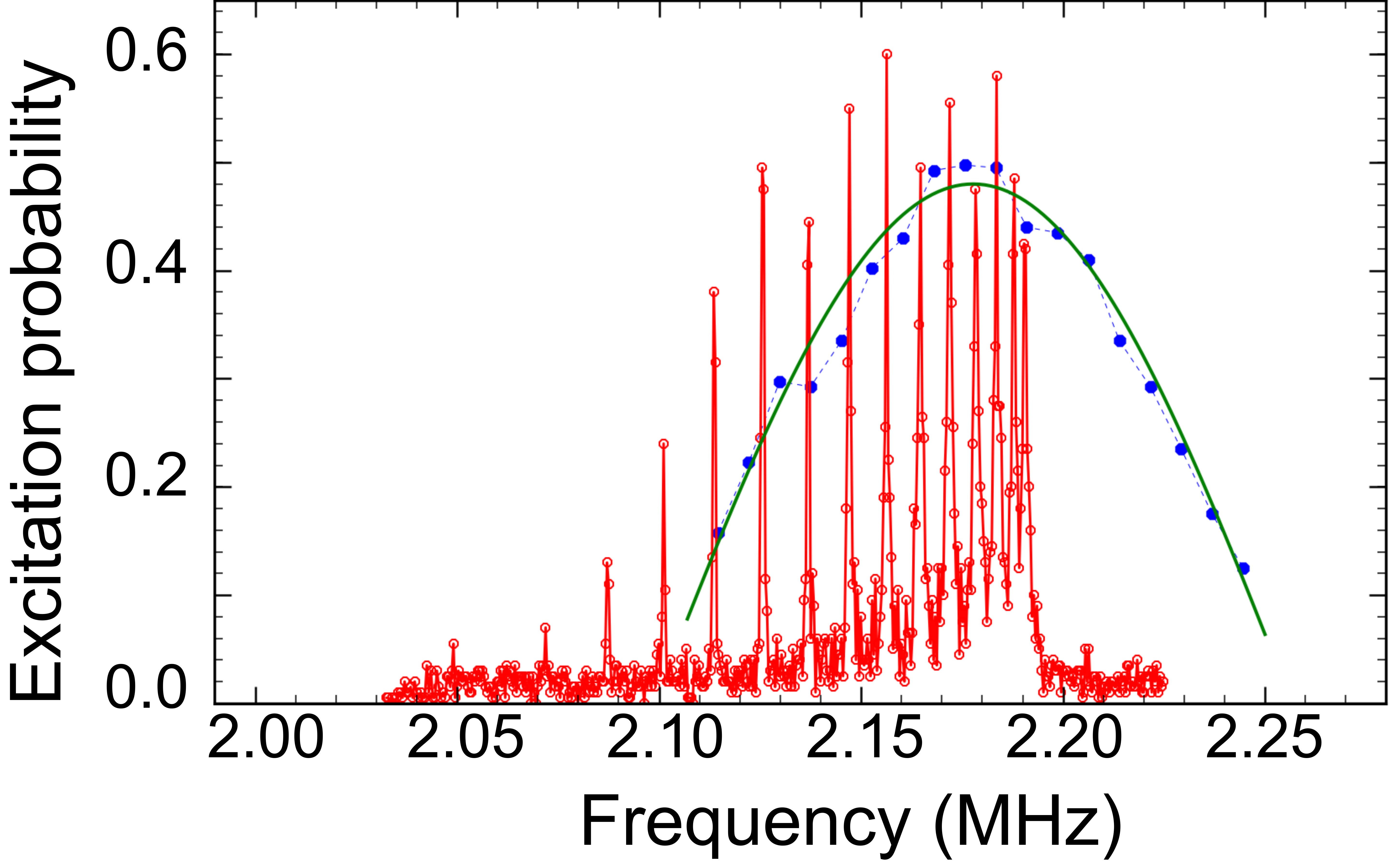}
\caption{ \label{fig:spectrum} \textbf{Spectra of blue transverse sidebands of a 42 ion string.} \new{Excitation probability to the 
$|D_{1/2}, m_j = -1/2 \rangle$ state} of the probe ion as a function of the detuning. 
Red: excitation time of $900\ ${\textmu}s, together with a relatively low intensity resolves individual normal modes. Visible are the normal modes of the $x$ direction. Blue: a short high-intensity pulse of length $8\ ${\textmu}s excites superpositions of the normal modes. Green: fit of the excitation to a sinc-function with only the amplitude and center frequency as free parameters.
}
\end{figure}
\end{center}

The string is cooled by 
using laser light at 397~nm, red detuned 
with respect to the $S_{1/2} \leftrightarrow P_{1/2}$ transition, to the Doppler limit  characterized by a mean 
phonon number $\nbar\approx 5$ of the transverse modes. To 
control the quantum information in the string, we 
drive the quadrupole transition between the $\ket{S}=\ket{S_{1/2}, m_j = -1/2}$ and $\ket{D}=\ket{D_{5/2}, m_j = -1/2}$ states by a 
729~nm laser beam \cite{Naegerl2000}, as schematically 
shown in Fig. \ref{fig:schematics}.
This beam is tightly focused addressing  the probe ion at the end of the ion chain, with a small projection on the $z$ axis, and equal projections on the $x$ and $y$ axis. To create correlations between the electronic and motional degree of freedom,
a short laser pulse $R^+(\theta,\varphi)$ with length $\theta$ and phase $\varphi$ tuned to the $+\omega_x$ sideband of the quadrupole transition 
induces the coupling $\ket{S,n} \leftrightarrow \ket{D,n+1}$, where $n$ labels a local motional Fock state of the addressed ion \cite{Leibfried2003,Wineland1998}. 
To initially prepare a localized vibrational excitation, the pulse duration of $5-10\ ${\textmu}s needs to be much shorter than the inverse coupling $t_{12}$ 
of the local motion to the 
adjacent ion. Figure~\ref{fig:spectrum} shows a spectrum near the blue sideband of the $x$ and $y$ 
modes of a 42-ion string with axial frequency $\omega_z = 2\pi \times$ 72.14~kHz. For long excitation times of 
900~{\textmu}s (red trace in Fig.~\ref{fig:spectrum}), we resolve the normal modes of the 42 ion chain. 


However, a short excitation time of 8~{\textmu}s at about 250 times the
intensity used to resolve the normal mode spectrum excites superpositions of the normal modes (blue trace in Fig.~\ref{fig:spectrum}) corresponding to the desired local excitation of the ion string \cite{Ramm2014}.

After this initial state preparation, we allow the initially localized phonon to travel into the phonon bath of the warm ion string before a second probe pulse 
interrogates the coherence between the qubit and the vibrational motion 
(see Fig.~\ref{fig:schematics}) \cite{Gessner2013b}. Full interference contrast can only be restored if the original vibrational excitation refocuses at the 
probe ion simultaneously with the probe pulse.  The observation of Ramsey fringes as a function of the relative phase between the two pulses can furthermore be traced back to qubit-phonon discord of the probe ion \cite{Gessner2013b,gessner2016local} (see Supplementary \new{Notes 1,2,3}).

\new{\subsubsection*{Theretical interpretation}}
For the probe ion  initially in the state $\ket{S,n}$, the 
preparation pulse $R^+(\theta,0)$ of length $\theta$ and phase 0 creates a superposition of the form $\alpha_n \ket{S,n}+ \beta_n \ket{D,n+1}$, where 
$\alpha_n$ and $\beta_n$ are specific for each Fock state $\ket{n}$.
The inverse operation can be applied with a 
second pulse, with the same parameters but opposite phase $\Delta \varphi = \pi$ with respect to the first pulse.

As long as the system has not evolved between both pulses, we expect full contrast. However, if the phonon created 
by the preparation pulse is no longer localized at the probe ion, the second pulse (see Fig.~\ref{fig:schematics}) cannot map the qubit-phonon coherences back to the electronic populations and therefore no phase dependency can be observed. Thus, loss of phase contrast indicates delocalization of the phonon excitation that is tagged by  the 
probe ion's internal state (see Supplementary \new{Notes 1,2,3} for a detailed derivation).

More specifically, the phase contrast, or visibility
\begin{eqnarray} \label{visibility}
v(t)=\frac{\max_\varphi(P_D(t)-\min_\varphi(P_D(t))}{\max_\varphi(P_D(t)+\min_\varphi(P_D(t))}\:,
\end{eqnarray} where $P_D$ is the probability that after the Ramsey sequence with free evolution time $t$ the ion is found in the excited $|D_{1/2}, m_j = -1/2 \rangle$ state. The visibility $v(t)$ is 
closely connected to both the autocorrelation function of the probe ion's vibrational degree of freedom
and to the discord 
between the electronic and the motional degrees of freedom of the probe ion (see Eqs. (19,29,32) of Supplementary Notes \new{1 and 3}). The reason for this is that all three quantities relate to the coherences between the electronic states $\ket{S}$ and $\ket{D}$ in the density matrix pertaining to the motional degrees of freedom of the probe ion. 

In order to make a quantitative connection, we assume that the density matrix describing the ion string after laser cooling is diagonal in the collective mode basis and that all modes are equally populated. This is justified by noting that  laser cooling acts on the individual ions on time scales faster than the coupling between ions and that the spread of the eigenmode frequencies are nearly degenerate.

With these assumptions, we find in Supplementary Note \new{2} (see Eqs.~19,22 therein) for the modulus of the
correlation function  $|C(t)|$
\begin{align}\label{corr-func}
 |C(t)|=|\langle a_1(t)a_1^{\dagger}(0)\rangle| = (\bar{n}+1) v(t),
\end{align}
where $\bar{n}=\langle a_i^\dagger a_i \rangle \approx 5$ is the average number of phonons in each of the normal modes. 
Further, we arrive at a simple measure for the quantum discord (Eq.~32 of Supplementary Note \new{3}):
\begin{eqnarray} \label{discord}
D(t) &=& \frac{\pi}{4} \frac{|C(t)|}{\bar{n}+1} = \frac{\pi}{4} v(t)\:.
\end{eqnarray}
To derive
this relation we have performed a perturbation expansion
for small $\theta/2 \approx 0.3$, including contributions of first and
second order, and used the additional assumption that the
initial phonon state may be approximated by an equilibrium
state of the local phonon modes (for details see Supplementary Note \new{3}.

\begin{center}\begin{figure*}
\includegraphics[width=0.65\textwidth]{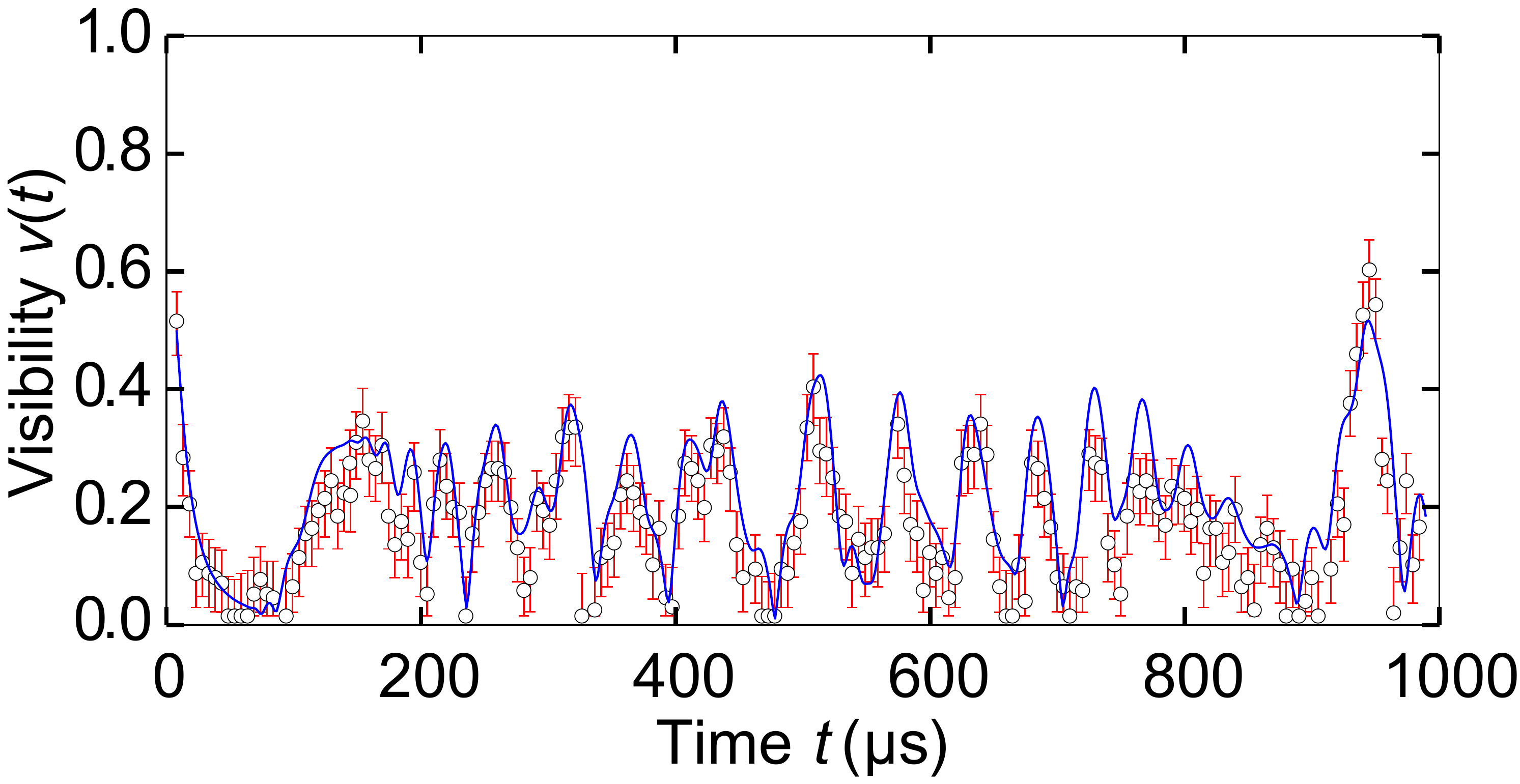}
\caption{ \label{fig:42-ions} \textbf{Visibility measurement for 42 ions.} Visibility (according to Eq.~\ref{visibility}) deduced from the population evolution of the state
$|D_{1/2}, m_j = -1/2 \rangle$ as a function of the free evolution time $t$ of the Ramsey sequence. 
A partial revival of the initial state population occurs at the rephasing times of the eigenfrequencies. Experimental results (red 
circles) are shown along with  theory (blue line), for a chain of 42 ions with an axial trapping frequency 
$\omega_z$ = 2$\pi \times$106.9~kHz. Error bars represent the Bayesian 90\% credible interval for the visibility. The only free parameter in the fit is an overall scale factor of the visibility of 0.67 to take into account loss of coherence mainly due to the incoherent background of the laser light (see text). 
}
\end{figure*}
\end{center}

\new{\subsubsection*{Experimental results}}
For the experiments, we first determine the time it takes to excite the 
probe 
ion to the $\ket{D}$-level with a probability of $0.5$, nearly saturating the transition for finite-temperature ion 
strings. We call this time the effective 
$\pi$-time $t_{\pi}$. The actual sequence to probe the dynamics of the crystal 
consists of two pulses each of length $t_{\pi}/2$ (see Fig. \ref{fig:schematics}) and separated by time $t$. To detect and quantify the discord between qubit and local
vibrational degree of freedom, we vary the phase difference $\Delta \varphi$ between the 
first and the second Ramsey pulses. In particular, we 
choose $\Delta \varphi = \{0, \pi/2, \pi, 3\pi/2$\} and extract the phase contrast $v(t)$ for each value of the free evolution time 
$t$. Figure \ref{fig:42-ions} shows a comparison of the experimentally inferred phase contrast for 42 ions 
to the theoretical expectations as given by Eq.~(19) in Supplementary Note \new{1}.

\begin{center}\begin{figure*}
\includegraphics[angle =0, width=0.99\textwidth]{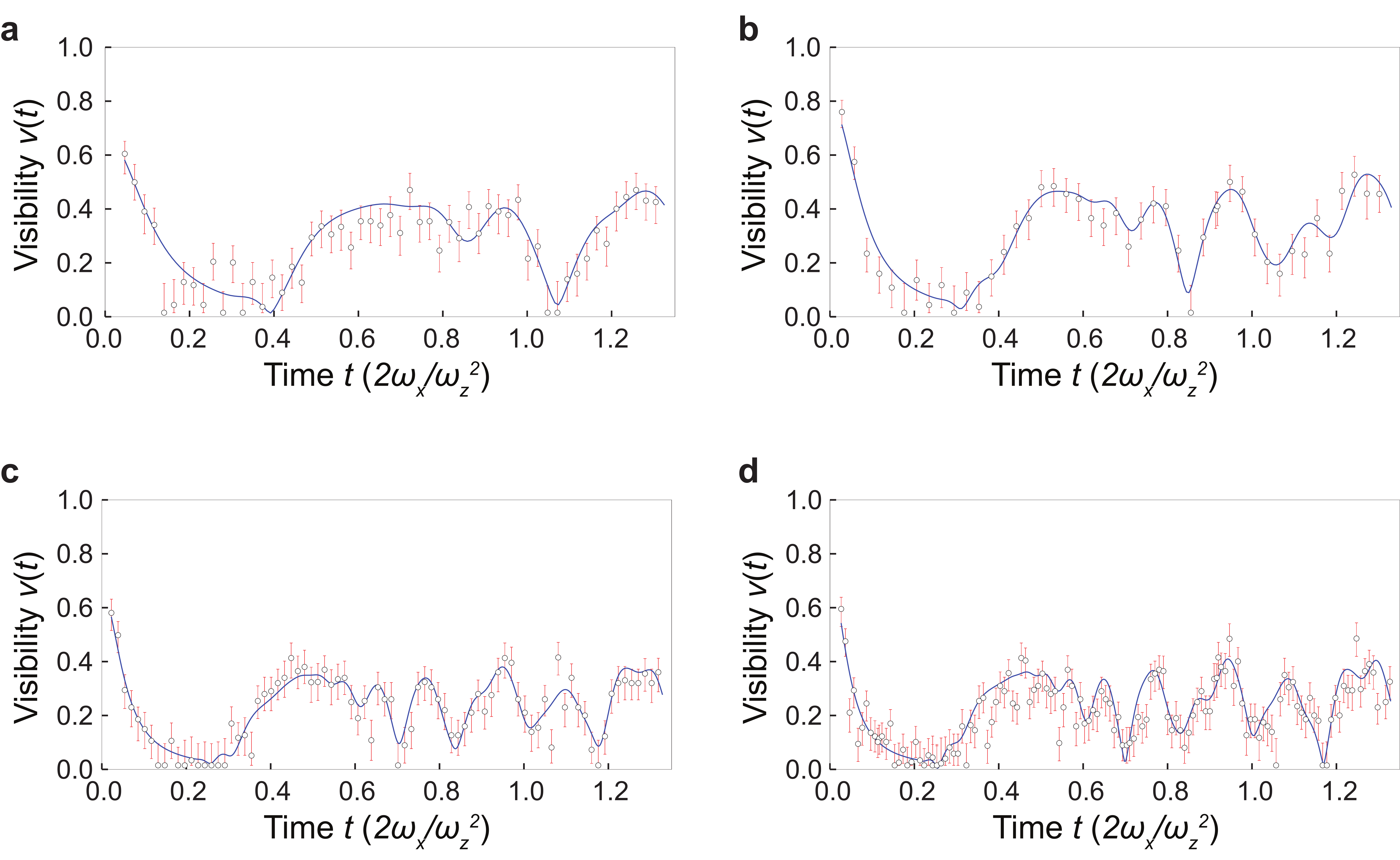}
\caption{\label{fig:more-ion-numbers} \textbf{Visibility measurements for \new{a variable number of ions}.}
Experimental result in \textbf{a,} is for 8 ions, \textbf{b,} 14 ions, \textbf{c,} 25 ions,
and \textbf{d,} with 33 ions where the axial trapping frequency takes the values
$\omega_z$ = 2$\pi \times$ (195.8~kHz, 155.2~kHz, 93.8~kHz, and 117.1~kHz), respectively.  \new{ Error bars represent the Bayesian 90\% credible interval for the visibility.} While the traces are very similar, increasing the number of ions tends to produce sharper features.
}
\end{figure*}
\end{center}

To gather sufficient statistics, we repeat the experiment 500-750 times per data point. For short times $t$, we observe a large phase contrast as the phonon excitation is still localized at the site of the 
probe ion. After a few tens of microseconds, the phonon excitation couples to the other sites \cite{Ramm2014} and the phase contrast diminishes. However, the phase contrast revives when the phonon excitation recurs at
the original site. This revival of the phase contrast also proves that the phase coherence of the initially injected phonon is maintained even 
as it delocalizes over the ion string. 
We also study this dynamics for 8, 14, 25 and 33 ions (see Fig.~\ref{fig:more-ion-numbers}). 

Common to all measurements is a rapid loss of phase 
contrast followed by a specific revival pattern.
Comparison with theory shows that the revival pattern is governed by the phonon dynamics in the ion chain. However, we also observe a reduction of the maximally expected visibility from 1 to values between 0.66 and 0.80. The reduction is independent of the Ramsey gap time, consistent with the measured Gaussian qubit decoherence with $T_2 \sim 2~{\rm ms}$. Instead, we attribute the reduced contrast mainly to a broad incoherent background of the laser light (see Methods).

\new{\subsection*{Discussion}}

We expect that suitable extensions of the 
scheme to protocols consisting of more than one ion or two pulses are in fact capable of extracting higher-order phonon correlation functions 
both in space and time. This way, the methods of non-linear spectroscopy, which are typically employed to study dynamical and spectral features of 
molecular aggregates and semiconductors \cite{Mukamel1995}, become accessible to probe quantum optical many-body systems. This opens up a powerful way of 
analyzing complex interacting quantum systems by measuring space and time correlations \cite{Gessner2014a,Schlawin2014}

In this context, it is also interesting to note that the method tracks an excitation in a bath of substantial size. In the example of a 42-ion string cooled down to the Doppler limit of about five local transverse quanta, there are about $42 \times 5 \approx 200$ phonons present. Our method generates a single phonon and tags it with a specific phase relationship to the electronic state of 
the probe ion, allowing us to re-identify this excitation when it returns to 
its origin. Hence, the method allows to follow the dynamics of single phonons in a finite temperature environment. 

Our measurements also show that quantum coherence of motional excitations can be preserved even in long ion strings. Direct extensions of this work are to measure how phonons scatter off impurities and how this affects the transport dynamics in a finite temperature environment\new{\cite{Cormick2016}}. This could be done by coupling some of the phonons to individual qubits of other ions via sideband interactions. These interactions implement a non-linear dynamics in the photon bath allowing one to study more complex cases than the linear dynamics studied here. Further extensions include perturbing  the potential with an optical lattice enabling studies of nanofriction and the Aubry-transition \cite{Bylinskii2015} or the interplay of disorder and interactions \cite{Basko2006}. 

More generally, it would interesting to study how this method could be applied to other systems, e.g., interacting spin chains, where individual excitations could be coherently traced via an auxiliary probe spin. 

Finally, our method is scalable in the sense that the control of the subsystem is independent of the size of the total system. Thus, the discussed method can be applied or adapted to large coherent systems where suitable single particle control is available, such as for instance for neutral atoms in optical lattices \cite{Bakr2009}.

\section{Acknowledgments}
This work has been supported by AFOSR through the ARO Grant No. FA9550-11-1-0318 and by the NSF CAREER programme grant no. PHY 0955650.
A.B. \& H.P.B. acknowledge support through the EU Collaborative Project QuProCS (Grant agreement 641277).

\new{
\section{Contributions}
H.H., M.G., M.R., and T.P. conceived the experiment. 
A.A., M.R., T.P., R.M., D.G., and P.S. carried out the measurements. 
O.K., M.G., H.P.B., and A.B. carried out the theoretical analysis, A.A., H.H., O.K., M.G., H.P.B., and A.B. wrote the manuscript. All authors contributed to the discussions of the results and manuscript.
}
\new{\section{Competing financial interests}
The authors declare no competing financial interests.}

\bibliographystyle{unsrt}

\clearpage

\section*{Methods}

\subsection*{Decoherence due to laser light}

We observe a reduction of the maximally expected visibility from 1 to between 0.66 and 0.80. The reduced contrast is mainly due to a broad incoherent background on the laser light on the order of $B=-40$~dB below the carrier of the laser light, however, peaking to around $B=$-28~dB near the bandwidth of 700~kHz of our servo loop to stabilize the laser frequency. This incoherent background drives other transitions than the blue sideband, and in particular the strong atomic carrier transitions between the S$_{1/2}\: (m_J=-1/2)$ and D$_{5/2}\: (m_J=\{-5/2,-3/2,-1/2, 1/2, 3/2\})$ levels.  Our geometry was not optimized to drive only the S$_{1/2}\: (m_J=-1/2)$ and D$_{5/2}\; (m_J=-1/2)$ transition, and hence we assume that all carrier transitions have a similar coupling strength of $\Omega_0\approx 150$~kHz when driving effective $\pi$-pulses on the blue sideband in $t_{\pi}=20\:${\textmu}s. Thus, the population of the D-state will increase by $\sin\left(\sqrt{B}\Omega_0 t_{\pi}/2\right)$, corresponding to an increase between 0.016~($B=-40$~dB) and 0.06 ($B=-28$~dB) for each $\pi_{\rm eff}/2$-pulse and transition. We estimate that this effect accounts for 0.2 of the loss of contrast for both pulses and all five atomic carrier transitions.

Another concern is motional heating during the free evolution time. 
In our trap, motional heating almost exclusively stems from voltage noise at the trap electrodes resulting in common noise to all ions. 
Thus, the center-of-mass mode heats with about 0.2 quanta / ms while all other modes heat with less than 0.01 quanta/ms. 
Because the heating is predominantly common mode, motional heating will not influence the intrinsic dynamics of the ion string. 

\new{
\section*{Data Availability}
All relevant data is available from the authors.
}

\clearpage

\setcounter{section}{0}

\renewcommand*\thesection{{SUPPLEMENTARY NOTE} \arabic{section}}

\section{Ramsey sequence on the local sideband transition}
The experimental sequence consists of two fast, focussed pulses, which are separated by free evolution, and each of which addressing the blue sideband transition of the outermost ion. The second local pulse is applied with a phase shift of $\pi$ with respect to the first pulse. In the following we describe the elementary ingredients for the resulting Ramsey scheme.

\subsection{Hamiltonian in the absence of laser pulses}
In the absence of laser pulses, the ion chain is described by the Hamiltonian
\begin{equation}
\label{eq:free}
H_{\rm free} = \hbar \omega_0 \sigma_+ \sigma_- + H_{\rm ph},
\end{equation}
where $\omega_0$ is the electronic (carrier) transition frequency and 
$\sigma_+ = \sigma_-^\dagger = \ket{D}\bra{S}$, with $\ket{ D}$ and 
$\ket{ S}$ denoting the electronic ground state $D$ and excited state $S$ 
of the outermost ion in the chain. Since other electronic levels as well as 
the electronic states of other ions are not populated in the course of the 
experimental sequence, they can be ignored in the theoretical description. 
The second term $H_{\rm ph}$ describes the motional states (phonons) of the 
chain as defined in Eq.~(1) of the main manuscript. The phonon Hamiltonian
$H_{\rm ph}$ is diagonalized in terms of annihilation operators $A_{ j}$ of the collective modes as
\begin{equation}
H_{\rm ph} = \sum_j \hbar \omega_j A_{ j}^\dagger A_{ j},
\end{equation}
where $\omega_j$ is the frequency of the $j$th eigenmode. The basis transformation between collective phonon modes $(A_{ j})$ and local modes $(a_i)$ is described by a unitary matrix of coefficients $\beta_{ij}$, such that
\begin{equation}\label{collective-ann}
 a_i = \sum_{\rm j=1}^{N} \beta_{ ij} A_{ j}.
\end{equation}

\subsection{Initial state}
After Doppler cooling and optical pumping, the ion chain is initialized in the quantum state:
\begin{equation}
\rho_{\rm in} = \ket{S}\bra{S} \otimes \rho_{\rm ph},
\end{equation}
where $\ket{ S}\bra{ S}$ denotes the electronic ground state of the first ion and $\rho_{\rm ph}$ represents a thermal state of motion, which is diagonal in the energy eigenbasis \cite{Morigi2003}
\begin{equation}\label{ph-A}
    \rho_{\rm ph} = \frac{1}{Z_A} e^{- H_{\rm ph}/k_{\rm B}T} ,
\end{equation}
with $Z_A=\tr\left( e^{-H_{\rm ph}/k_{\rm B}T}\right)$, Boltzmann constant $k_{\rm B}$ and temperature $T$. 

\subsection{Sideband pulses}
The Ramsey pulses in this experiment couple the electronic states of the first ion, $\ket{S}$ and $\ket{D}$, to its motional states via sideband transitions. In the experiment, both Ramsey pulses are focused on the same single ion, while being applied with a pulse duration of $5-13$~{\textmu}s, which leads to a spectral spread of $2\pi \times (69-178)\mathrm{kHz}$. In all of these experiments the spread of the eigenmodes that are involved in the motion of the first ion (with Lamb-Dicke factors of larger than 0.01) is less than $2\pi \times 58 \mathrm{kHz}$. 
Therefore, the pulse occurs on a faster time-scale than the phonon tunneling, thereby creating localized phonons at the first ion. In a suitable interaction picture, this local sideband transition is therefore described by the Hamiltonian
\begin{equation}\label{eq:blue-op-exact}
H =   \frac{\hbar\Omega_0}{2} \left(\sigma_+ e^{i( \mathbf{k}_L\cdot\mathbf{x}_1(t)-\Delta \cdot  t + \phi)} + \sigma_- e^{-i( \mathbf{k}_L\cdot\mathbf{x}_1(t) -\Delta \cdot t + \phi)} \right), 
\end{equation}
in which $\mathbf{x}_1(t) = U_{\rm free}(t)^\dagger \mathbf{x}_1 U_{\rm free}(t)$, where $\mathbf{x}_1$ is the position operator of the first ion, $\mathbf{k}_L$ is the wavevector of the laser, $\Omega_0$ is the Rabi frequency of the driving laser with frequency $\omega_L = \omega_0 + \Delta$ and phase $\phi$. Here, $\Delta$ is chosen as the frequency of radial oscillations of the first ion 
along the direction of $\mathbf{k}_L$.

The ion chain is cooled to the Doppler limit to a mean phonon occupation number of $\nbar \approx 5$. With a Lamb-Dicke parameter of ${\eta \approx 0.05}$, we obtain $\eta \sqrt{\bar{n}} \approx 0.11\ll 1$, hence, we can approximate the Hamiltonian further by assuming the Lamb-Dicke regime \cite{Leibfried2003}. Moreover, we apply a rotating-wave approximation (discarding fast-rotating terms that contain the operators $a_1^{\dagger}\sigma_+$ and $a_1\sigma_-$) to finally obtain the unitary
operator describing the action of the first Ramsey pulse
\begin{equation}\label{eq:blue-op}
R(\theta, \phi) = \exp\left\{  \frac{\theta}{2} \left(e^{i\phi} \sigma_+ a_1^{\dagger} - e^{-i\phi} \sigma_- a_1\right) \right\}, 
\end{equation}
in which $\theta= \eta \Omega_0 \tau$ and $\tau$ denotes the pulse duration.

\subsection{Time-evolution under the Ramsey pulse sequence}
Let us now describe the evolution of the initial quantum state $\rho_{\rm in}$ under the Ramsey sequence. We first focus on the description of the quantum state after an initial sideband pulse with phase $\phi_1$, followed by free evolution for a time $t$, which leads to the following quantum state:
\begin{equation}
    \rho(\theta, \phi_1; t)  = U_{\rm free}(t) R(\theta, \phi_1) \rho_{\rm in} R^\dagger(\theta, \phi_1) U^\dagger_{\rm free}(t).
\end{equation}
The free evolution is described by $U_{\rm free}(t)=\exp\left(-iH_{\rm free}t\right)$, where $H_{\rm free}$ was introduced in Eq.~(\ref{eq:free}). Since $[H_{\rm free},\rho_{\rm in}]=0$, we can rewrite the above expression as
\begin{equation}\label{eq.dmfirstpulsefree}
    \rho(\theta, \phi_1; t)  =  R(\theta, \phi_1;-t) \rho_{\rm in} R^\dagger(\theta, \phi_1, -t), 
\end{equation} 
where $R(\theta, \phi_{i}; -t) =U_{\rm free}(t) R(\theta, \phi_{i}) U^\dagger_{\rm free}(t)$. We further express $R(\theta, \phi; -t)$ in terms of collective modes,
\begin{equation}
\begin{aligned}
R(\theta, \phi; -t) &= \exp\left(-i \sum_{i=1}^{N} \omega_{i} A^\dagger_{i} A_{i} t \right) \exp\left\{  \frac{\theta}{2} \left(e^{i(\phi-\omega_0t)} \sigma_+ \sum_{j=1}^{N} \beta_{1j}^* A_{ j}^{\dagger} - e^{-i(\phi-\omega_0t)} \sigma_- \sum_{j=1}^{N}\beta_{1j} A_{j}\right) \right\}\notag\\&\quad\times\exp\left(i \sum_{i=1}^{N} \omega_{i} A^\dagger_{i} A_{i} t \right).
\end{aligned}
\end{equation}
The bosonic commutation relations of the collective mode operators lead to interaction-picture operators of the form $e^{-i \omega_j A_j^\dagger A_j t} f(A_j^\dagger, A_j) e^{i \omega_j A_j^\dagger A_j t} = f(A_j^\dagger e^{-i\omega_jt}, A_j e^{i\omega_jt})$, where $f(A_j^\dagger, A_j)$ is some function of the collective mode operators $A_j^\dagger$ and $A_j$. We write
\begin{equation}
R(\theta, \phi; -t) = \exp \left(\frac{\theta}{2} u(\phi,-t)\right) ,
\end{equation} 
where $u(\phi,-t) = e^{i(\phi-\omega_0t)} \sigma_+ A(t)^{\dagger} - e^{-(i\phi-\omega_0t)} \sigma_- A(t)$. We have further introduced 
\begin{align}\label{eq.Afunction}
A^\dagger(t) = a_{1}^\dagger(-t) =\sum_{j=1}^{N} \beta_{1j}^* A_j^\dagger e^{-i \omega_j t}= \sum_{k=1}^{N} \gamma_{ k}(t) a_k^\dagger,
\end{align}
with 
\begin{equation}\label{gamma}
\gamma_{k}(t) = \sum_{j=1}^{N} \beta_{1j}^* \beta_{kj} e^{-i \omega_{j} t}.
\end{equation}
Finally, after application of the second Ramsey pulse (with phase $\phi_2$ and same duration $\theta/2$ as the first one), the density matrix is given by
\begin{align}\label{eq.finalstate}
    \rho_{\rm f}(\theta, \phi_1,\phi_2; t)  &=   R(\theta, \phi_2;0)  \rho(\theta, \phi_1;t)R^\dagger(\theta, \phi_2;0), \notag\\
&= R(\theta, \phi_2; 0) R(\theta, \phi_1; -t)\rho_{\rm in}R^\dagger(\theta, \phi_1; -t)R^\dagger(\theta, \phi_2; 0).
\end{align} 

\subsection{Visibility}
In the experiment, we measure the population of the $| D\rangle$ state, which is given by
\begin{equation}\label{sib}
P_{D}(\theta, \phi_1,\phi_2; t) = \tr \left [P_{ D} \rho(\theta, \phi_1, \phi_2; t) \right ],
\end{equation}
where $\tr$ denotes the trace, $P_{ D}=\ket{ D}\bra{ D}\otimes\mathbb{I}_N$, and $\mathbb{I}_N$ is the identity operator on the Fock space of the $N$ bosonic modes. With the final state $\rho_{\rm f}(\theta, \phi_1,\phi_2; t)$, as defined in Eq.~(\ref{eq.finalstate}), we obtain
\begin{equation}\label{eq:PD}
P_{D}(\theta, \phi_1, \phi_2; t) = \tr \left[P_{ D}\exp \left(\frac{\theta}{2} u(\phi_2,0)\right) \exp \left(\frac{\theta}{2} u(\phi_1,-t)\right) \rho_{\rm in} \exp \left(-\frac{\theta}{2} u(\phi_1,-t)\right) \exp \left(-\frac{\theta}{2} u(\phi_2,0)\right)  \right ]. 
\end{equation}
This result is valid in the Lamb-Dicke regime for arbitrary pulse duration $\theta/2$. To find an analytical expression for the Ramsey fringe visibility we approximate this expression further, based on experimentally motivated assumptions.

In the experiment, the normal mode frequencies are of the order of $\omega_j\approx 2 \pi \times 2\,$MHz, while the spread of normal modes is given by $\Delta\omega \lesssim 2 \pi \times 100\,$kHz. Therefore, $\Delta \omega/\omega_j \approx 0.05$, which implies that the average number of phonons in normal modes $\bar{n}_{j} = \langle A_{ j}^\dagger A_{ j} \rangle_{\rho_{\rm ph}}=\tr\{A_{ j}^\dagger A_{ j}\rho_{\rm ph}\}$ are approximately equal. Hence, we assume that the initial state describes an incoherent, homogeneous phonon distribution among the collective modes, i.e.,
\begin{equation}\label{eq:collective-modes}
\left< A_{ j}^\dagger A_{ k} \right>_{\rho_{\rm ph}} \approx \bar{n}\delta_{jk}.
\end{equation}

Furthermore, we expand $R(\theta, \phi; -t)$ up to second order in the pulse duration $\theta/2$. In all of our measurements, $\theta/2\lesssim 0.32$. We obtain
\begin{equation}
P_{D}(\theta, \phi_1, \phi_2; t) = \left(\frac{\theta}{2}\right)^2 \left(\bar{n} + 1 \right)\sum_{j,k,l=1}^{N} \left(  e^{i\phi_2}\gamma_{j}(t)+e^{i\phi_1}\delta_{j1} \right) \left(  e^{-i\phi_2}\gamma_{j}^*(t)+e^{-i\phi_1}\delta_{ j1} \right) \beta_{jk}^{*} \beta_{lk} 
 +  \mathcal{O}(\theta^4 ),
\end{equation}
where $\bar{n}$ is the average phonon occupation number per normal modes. Using $\sum_{j=1}^{N}|\gamma_{ j}(t)|^2=1$ and $\Delta\phi = \phi_1-\phi_2$, the expression above can be simplified to  
\begin{equation}
P_{D}(\theta, \phi_1, \phi_2; t) = \left(\frac{\theta}{2}\right)^2 \left( \bar{n} + 1\right) \left( 2 + e^{-i(\Delta\phi)} \gamma_1(t) +  e^{i(\Delta\phi)} \gamma_1^*(t)\right)  +  \mathcal{O}(\theta^4 ).
\end{equation}
The fringe visibility is now given by
\begin{equation}\label{eq:vis}
v(t) = \frac{\max(P_{D}(t)) - \min(P_{D}(t))}{\max(P_{D}(t))+\min(P_{D}(t))}\approx |\gamma_1(t)|,
\end{equation}
where the maximum and minimum are taken by optimizing over the phase difference $\Delta\phi$ between the two Ramsey pulses. The final result is obtained by writing $\gamma_1(t)=|\gamma_1(t)| e^{i \varphi(t)}$ and realizing that $\max(P_{D}(t))$ and $\min(P_{D}(t))$ are attained at $\Delta\phi = \varphi(t)$ and $\Delta\phi = \varphi(t)+\pi$, respectively. The function $\gamma_1(t)$ was introduced in Eq.~(\ref{gamma}). The result~(\ref{eq:vis}) was used to predict the time evolution of the visibility in Figures 3 and 4 of the main manuscript.


\section{Two-point correlation function}
The above Ramsey scheme can be interpreted as a method to extract the two-point auto-correlation function of the phonons by means of spin measurements. To see this, consider the first ion's phonon annihilation operator in the Heisenberg picture, which using Eq.~(\ref{collective-ann}) reads,
\begin{equation}
 a_1(t) = e^{+iH_{\rm free}t} a_1 e^{-iH_{\rm free}t} = 
 \sum_j \beta_{1j} A_j e^{-i\omega_jt}.
\end{equation}
With this, the two-point auto-correlation function of the phonons is given by
\begin{equation}
 \langle a_1(t)a_1^{\dagger}(0)\rangle 
 = \sum_{jk} \beta_{1j}\beta_{1k}^{\ast} e^{-i\omega_jt} \langle A_j A^{\dagger}_k \rangle.
\end{equation}
With the assumption of an initial phonon state that is diagonal in the collective mode basis, and whose population is independent of the mode index $j$, c.f. Eq.~(\ref{eq:collective-modes}), we obtain
\begin{align}\label{corr-func}
 \langle a_1(t)a_1^{\dagger}(0)\rangle 
 &= \sum_{j} \beta_{1j}\beta_{1j}^{\ast} e^{-i\omega_jt} \langle A_j A^{\dagger}_j \rangle\notag\\
 &= (\bar{n}+1) \gamma_1(t),
\end{align}
where, again, $\bar{n}=\langle A_j^{\dagger} A_j \rangle$. Thus, the experimental scheme may be seen as a local method to directly determine the modulus of the two-point correlations functions of the phonon chain. 

Based on this argument, we expect that suitable extensions of the scheme to protocols consisting of more than two pulses are in fact capable of extracting also higher-order phonon auto-correlation functions. This way, the methods of non-linear spectroscopy, which are typically employed to study dynamical and spectral features of molecular aggregates and semiconductors \cite{ShaulBook}, become accessible to probe this many-body system. This opens up a powerful way of analyzing complex interacting quantum systems, especially if combined with the single-site addressability that is unique to quantum optical systems \cite{NJP14,SGMB14,JCP15,Lemmer}.

\section{Monitoring of spin-phonon discord dynamics}
In this section, we show that the Ramsey signal discussed above reveals the dynamics of discord-type correlations between the electronic and motional degree of freedom of the first ion. To this end, we revisit the density matrix $\rho(\theta, \phi_1; t)$ of the total system after the first sideband pulse and after free evolution of the chain (duration $t$), which was previously introduced in Eq.~(\ref{eq.dmfirstpulsefree}). Within second order in $\theta/2$ the density matrix is given by
\begin{align} \label{rho-total}
\rho(\theta, \phi_1; t) &=
 |0\rangle\langle 0| \otimes \Big[ \rho_{\rm ph} - \frac{\theta^2}{8}
 \Big( A(t)A^{\dagger}(t) \rho^{(1)}_{\rm ph} + \rho_{\rm ph} A(t)A^{\dagger}(t) \Big) \Big]
 \nonumber \\
 &\quad + |1\rangle\langle 1| \otimes \frac{\theta^2}{4} A^{\dagger}(t) \rho_{\rm ph} A(t) 
 \nonumber \\
 &\quad + \frac{\theta}{2} \Big(
 |1\rangle\langle 0| \otimes e^{i\phi_1} A^{\dagger}(t) \rho_{\rm ph} +
 |0\rangle\langle 1| \otimes e^{-i\phi_1} \rho_{\rm ph} A(t) \Big).
 \end{align}
where $A(t)$ was introduced in Eq.~(\ref{eq.Afunction}). Taking the trace over the local phonons $i=2,\ldots,N$ one finds the density matrix describing the 
combined quantum state of the electronic and the motional degrees of freedom of the first ion:
\begin{align} \label{rho-phonon-1}
 \rho^{(1)}(t,\tau_1,\phi_1) &= \tr_{[2,N]} \rho(t,\tau_1,\phi_1) \nonumber \\
 &= |0\rangle\langle 0| \otimes Y(t) +  |1\rangle\langle 1| \otimes Z(t) \\
 &\quad + \frac{\theta}{2} \Big(
 |1\rangle\langle 0| \otimes e^{i\phi_1} \gamma_1(t) a_1^{\dagger}\rho^{(1)}_{\rm ph} +
 |0\rangle\langle 1| \otimes e^{-i\phi_1} \gamma^{\ast}_1(t) \rho^{(1)}_{\rm ph} a_1 \Big).
 \nonumber
 \end{align}
Here, $\rho^{(1)}_{\rm ph}=\tr_{[2,N]}\rho_{\rm ph}$ is the reduced state of the motional 
degree of freedom of the first ion, while $Y(t)$ and $Z(t)$ are certain operators acting on the 
corresponding state space. In order to derive (\ref{rho-phonon-1}) from (\ref{rho-total}) we have 
used the relation $\tr_{[2,N]} \{\rho_{\rm ph} a_k \} = \delta_{k1}\rho^{(1)}_{\rm ph}a_1$.

With the help of expression (\ref{rho-phonon-1}) we can easily determine the quantum discord between the electronic and the motional degrees of freedom of the first ion. A simple way to estimate the discord-type correlations is obtained from the dephasing-induced disturbance \cite{GBPRL11,GBPRA13,PhDGessner}, given by
\begin{equation}
 D(t) = \frac{1}{2} || \rho^{(1)}(t,\tau_1,\phi_1) - \rho^{(1)}_{\rm deph}(t,\tau_1,\phi_1) ||,
\end{equation}
where $||\cdot||$ denotes the trace norm and $\rho^{(1)}_{\rm deph}(t,\tau_1,\phi_1)$ is the
state obtained from the state $\rho^{(1)}(t,\tau_1,\phi_1)$ after dephasing in its eigenbasis:
\begin{equation}
 \rho^{(1)}_{\rm deph}(t,\tau_1,\phi_1) =
 |0\rangle\langle 0| \otimes Y(t) +  |1\rangle\langle 1| \otimes Z(t).
 \end{equation}
Thus, we have $D(t)=\frac{1}{2} ||X(t)||$, where
\begin{eqnarray}
 X(t) &\equiv& \rho^{(1)}(t,\tau_1,\phi_1) - \rho^{(1)}_{\rm deph}(t,\tau_1,\phi_1) \nonumber \\
  &=& \frac{\theta}{2} \Big(
 |1\rangle\langle 0| \otimes e^{i\phi_1} \gamma_1(t) a_1^{\dagger}\rho^{(1)}_{\rm ph} +
 |0\rangle\langle 1| \otimes e^{-i\phi_1} \gamma^{\ast}_1(t) \rho^{(1)}_{\rm ph} a_1 \Big).
 \nonumber
 \end{eqnarray}
To evaluate this expression analytically, we realize another approximation. In particular, we consider the local phonon state to be of the form
\begin{equation}
\rho^{(1)}_{\rm ph} =  \sum_{n_1=0}^{\infty} p_{n_1} |n_1\rangle\langle n_1|.
\end{equation}
A detailed discussion of this approximation will be given at the end of this section. Using this, one finds that the eigenvalues of $X(t)$ are given by 
\begin{equation}
 \pm\frac{\theta}{2}|\gamma_1(t)|p_{n_1}\sqrt{n_1+1}, \qquad n_1=0,1,2,\ldots,
\end{equation} 
which yields:
\begin{eqnarray} \label{discord}
 D(t) &=& \frac{1}{2} ||X(t)|| = \sum_{n_1=0}^{\infty} 
 \frac{\theta}{2}|\gamma_1(t)|p_{n_1}\sqrt{n_1+1} \nonumber \\
 &=& \frac{\theta}{2} \Big\langle \sqrt{n_1+1} \Big\rangle |\gamma_1(t)|.
\end{eqnarray}
This shows that the discord-type quantum correlations of the degrees of freedom 
of the first ion are proportional to the visibility $|\gamma_1(t)|$. The decay and revival of the 
visibility observed in the experiment thus correspond to a loss and gain of the quantum
correlations between the electronic and motional degrees of freedom of the first ion of the chain.

We can further combine Eqs.~(\ref{corr-func}) and (\ref{discord}), leading to:
\begin{equation}
 D(t) = \frac{\theta}{2} \frac{\Big\langle \sqrt{n_1+1} \Big\rangle}{\bar{n}+1}
 \Big|\langle a_1(t)a_1^{\dagger}(0)\rangle\Big|
 \approx 
 \frac{\theta}{2} \frac{\Big|\langle a_1(t)a_1^{\dagger}(0)\rangle\Big|}{\sqrt{\bar{n}+1}}.
\end{equation}
The measure of discord-type quantum correlations is thus directly proportional to the modulus of the two-point 
correlation function of the motional amplitude of the first ion. This result is valid 
for purturbatively small values of $\frac{\theta}{2}$. For the special case of $\pi/2$ pulses as applies to present experiments,

\begin{equation}
 D(t) \approx 
 \frac{\pi}{4} \frac{\Big|\langle a_1(t)a_1^{\dagger}(0)\rangle\Big|}{\bar{n}+1},
\end{equation}

in which we replaced $\theta\Big\langle\sqrt{n+1} \Big\rangle$ with $\pi/2$ indicating a Ramsey $\pi/2$-pulse from equation (\ref{eq:blue-op}). In this case, the perturbation parameter is $\frac{\theta}{2}\approx 0.3$. The equation above leads to 

\begin{equation}\label{discord-vis}
D(t) \approx \frac{\pi}{4} v(t),
\end{equation}
which indicates a direct proportionality between discord and visibility, and thereby proves visibility as a quantitative measure for discord-type quantum correlations in the chain.
 

To determine the coherences and the discord of the state of the electronic and the motional 
degrees of freedom of the first ion we have used the additional assumption that the phonon state 
(\ref{ph-A}) may be replaced by an equilibrium state of the local modes $a_j$:
\begin{equation} \label{ph-a}
 \rho^a_{\rm ph} = \frac{1}{Z_a} \exp \left[ -\beta \sum_j \hbar\omega_j a^{\dagger}_j a_j \right].
\end{equation}
Averages taken with $\rho^A_{\rm ph}$ are denoted by $\langle ... \rangle_A$ and averages
with $\rho^a_{\rm ph}$ are denoted by $\langle ... \rangle_a$. To justify the replacement
$\rho^A_{\rm ph}\rightarrow\rho^a_{\rm ph}$ we require that all first and second moments of
$A_j$ determined by $\rho^A_{\rm ph}$ and $\rho^a_{\rm ph}$
are identical. It is clear that we have the following exact relations:
\begin{eqnarray}
 \langle A_i \rangle_A &=& \langle A_i \rangle_a = 0, \\ 
 \langle A_i A_j \rangle_A &=& \langle A_i A_j \rangle_a = 0.
\end{eqnarray}
Thus, we only have to demand that also
\begin{equation}
 \langle A^{\dagger}_i A_j \rangle_A = \langle A^{\dagger}_i A_j \rangle_a
\end{equation}
holds, which yields the condition
\begin{equation}
 \delta_{ij} \bar{n}_j = \sum_{k} \beta_{ki}\beta^{\ast}_{kj} \bar{n}_k.
\end{equation}
By condition~(\ref{eq:collective-modes}) this becomes
\begin{equation}
 \delta_{ij} \bar{n} = \sum_{k} \beta_{ki}\beta^{\ast}_{kj} \bar{n},
\end{equation}
which is obviously satisfied since the matrix $(\beta_{ij})$ is unitary.

\section*{Supplementary References}
\bibliographystyle{plain}

\end{document}